\documentclass{elsarticle}

\usepackage{algorithmic}
\usepackage{algorithm}
   \usepackage[]{graphicx}
\usepackage{amsmath}
\usepackage{algorithmic}
\usepackage{url}

\begin{document}
%
\title{Defending against Co-residence Attack in Energy-Efficient Cloud: An Optimization based Real-time Secure VM Allocation Strategy}
%
%
%
%

\author{Lu~Cao,
        Ruiwen~Li,
        Xiaojun~Ruan,
        Yuhong~Liu}

\begin{abstract}

Resource sharing among users serves as the foundation of cloud computing, which,  however, may also cause vulnerabilities to diverse co-residence attacks launched by malicious virtual machines (VM) residing in the same physical server with the victim VMs. In this paper, we aim to defend against such co-residence attacks through a secure, workload-balanced, and energy-efficient VM allocation strategy. Specifically, we model the problem as an optimization problem by quantifying and minimizing three key factors: (1) the security risks, (2) the power consumption and (3) the unbalanced workloads among different physical servers. Furthermore, this work considers a realistic environmental setting by assuming a random number of VMs from different users arriving at random timings, which requires the optimization solution to be continuously evolving. As the optimization problem is NP-hard, we propose to first cluster VMs in time windows, and further adopt the Ant Colony Optimization (ACO) algorithm to identify the optimal allocation strategy for each time window. Comprehensive experimental results based on real world cloud traces validates the effectiveness of the proposed scheme.

\end{abstract}
\maketitle




%

\section{Introduction}\label{sec:introduction}

Cloud computing has become popular in both business and personal services. Infrastructure as a Service (IaaS) in cloud computing is a service model that grants multiple users' access to a shared pool of physical resources in a dynamic way. Such resource sharing allows the cloud to maximize the system efficiency by fully utilizing available computing resources. On the other hand, cloud users can dramatically save costs by paying only for the resources that they are using and releasing the idle resources to other users. These advantages attract numerous businesses that want to reduce costs on intensive computational operations. 

However, such infrastructure-level computing resource sharing, which is enabled through multi-tenancy (defined as ``the practice of placing multiple tenants on the same physical hardware"~\cite{tan2013multitenancy}), also introduces new security risks. Attackers taking advantage of the co-residence opportunities may perform diverse attacks against their co-tenants~\cite{standaert2010introduction, ristenpart2009hey,aciiccmez2007power, aciiccmez2006predicting,aciiccmez2007yet, grunwald2002microarchitectural, chiang2014swiper,varadarajan2012resource,momm2011combined, feng2011shrew}, threaten the security of cloud infrastructure and undermine users' confidence to move to the cloud~\cite{takabi2010security,subashini2011survey,ren2012security,chow2009controlling}. For example, a misconfigured hypervisor which hosts multiple Virtual Machines (VM) from different tenants may serve as a conduit for information leakage~\cite{PCI}. Chiang proposed Swiper attack with which the attacker uses a carefully designed workload to incur significant delays to the targeted co-resident application~\cite{chiang2014swiper}. Ristenpart and Swift proposed an attack which modifies the workload of a victim VM in a way that frees up resources for the attacker's VM~\cite{varadarajan2012resource}. Particularly, such co-residence attacks have two unique characteristics. (1) It is directly enabled by the resource sharing among different users, and will continuously exist unless users are isolated on different Physical Machines (PM). (2) It mainly leverages the legitimate resource requests. Therefore, conventional security techniques, such as authentication, authorization and access control, can hardly detect and block co-residence attacks without preventing normal access to the shared resources~\cite{varadarajan2012resource}.



There are a number of solutions proposed to defend against co-residence attacks through performance isolation which requires virtualized computing resource isolation for storage, CPU, cache, memory, and access path networks~\cite{martin2012timewarp,wu2012xenpump}. However, such solutions are typically either impractical (e.g., high overhead or nonstandard hardware), application-specific, or insufficient for fully mitigating the risk. Furthermore, it requires that the resources can never be overcommitted due to the possibility of concurrent requests from multiple tenants. This requirement will inevitably leave resources idle and sacrifice cloud performance and efficiency. Due to the immaturity of virtualization technology and the absence of physical isolation, smart adversaries are still able to launch attacks that penetrate the virtual boundaries among tenants \cite{chiang2014swiper, luo2011virtualization,jasti2010security,hyde2009survey,secVir,Minjiezheng}. At the current state of the art, there is no practical way to guarantee the unconditional security except avoiding multi-tenancy~\cite{ristenpart2009hey}.
 
Recently, a few studies have been proposed to focus on secure VM allocation strategies, which assign VMs to available physical machines (PMs) in a secured way to prevent malicious users from achieving co-residence with normal users~\cite{han2017using,azar2014co,han2013security}. Compared to performance isolation approaches, this type of mechanisms does not require significant changes of the existing hardware/software, and is not limited to specific applications. Nevertheless, this line of research has just been initiated recently and has very limited amount of work. In addition, as the number of possible allocations increases in a factorial way when the number of available PMs/VMs becomes large, it has been verified as an NP-hard problem to search for the best allocation~\cite{han2017using,han2013security}. Most of current studies resolve this issue only through heuristic solutions.

Therefore, a secure and energy-efficient VM allocation strategy to defend against the co-residence attacks is proposed in this paper. The main contributions of this research are summarized as follows.
\begin{itemize}
\item First, we propose to consider and quantify three key factors for secure VM allocation in energy-efficient cloud: (1) the security risks introduced by the co-residence of VMs from multiple users, (2) the overall power consumption and (3) the workload inequality among different PMs. The VM allocation problem is then modeled as an optimization problem where the objective function is to minimize these three factors at the same time. 

\item Second, this work assumes a realistic scenario where a random number of VMs from different users may arrive at the cloud end with random timings, which requires the optimization solution to be dynamically evolving based on both the existing allocation status and new allocation requests.  

\item Third, as this optimization problem is NP-hard \cite{han2017using}, we aim to address the problem by balancing the optimization goal, the computational complexity and the allocation delay. Specifically, we propose to first introduce time windows to handle arriving VMs in clusters. Then for each time window, the Ant Colony Optimization (ACO) algorithm, an evolutionary algorithm inspired by natural ant activities, is adopted to identify the optimal allocation strategy for new VMs based on the prior VM allocation status. Although ACO has already been applied to address diverse optimization problems, we are the first one to adopt it in the secure cloud resource allocation scenario. Comprehensive understanding and analysis on the physical meanings of (1) the ACO algorithm and (2) the cloud secure VM allocation scenario have been performed to facilitate such adoption. 

\item Fourth, comprehensive experiments based on real-world cloud workload traces are conducted to study (1) the impact of critical parameter settings; (2) the effectiveness of the proposed scheme when compared to the state-of-the-art secure VM allocation studies. 
\end{itemize}

\section{Background and Related Work}

\subsection{\bf Performance Isolation}
Diverse studies have been conducted to prevent sensitive information from being transferred through converted channels (i.e. side channels) between co-resident VMs at different levels of cloud infrastructure. First, eliminating side channels from hardware level \cite{wang2007new, keramidas2008non,martin2012timewarp} usually provides more effective defense. However, due to the complex process of introducing new hardware into existing cloud infrastructure, the adoption of such schemes adds extra cost on hardware and administration. Second, extensive researches have been carried out at the hypervisor level. For example, XenPump proposed as a module located in hypervisor \cite{wu2012xenpump}, monitors the hypercalls used by timing channels and adds latency to potential malicious operations, which increases the error rate in timing channels. In addition, Shacham et al. proposed to make the timer substantially more coarse by removing resolution clocks on Xen-virtualized x86 machines, so that malicious VMs can hardly obtain accurate time measurement~\cite{vattikonda2011eliminating}. The key drawback of these schemes is that they often require significant modifications on hypervisors. Third, some schemes are proposed at VM OS level \cite{zhang2013duppel} or application level \cite{coppens2009practical}. For instance, the authors in \cite{waheed2015security} proposed to hide real power consumption information from user VMs by deploying a police VM to generate false information. Such schemes do not require substantial changes in the cloud infrastructure and are thus easy to be adopted. Nevertheless, they often suffer from the heavy overhead caused by obfuscating side channel information at the upper level of the cloud infrastructure.


\vspace{-2mm}
\subsection{\bf Virtual Machine Allocation}
Attackers who aim to launch co-residence attacks against a certain target have to first place their malicious VMs on the same physical host where the target VM locates. Co-residence attacks cannot succeed if this first step fails. Therefore, researches are launched to design security aware VM allocation policies which significantly increase the difficulties for attackers to achieve co-residence.

Many VM allocation policies are studied to assign different positions to VMs. For instance, a randomization way to assign VMs has been proposed \cite{azar2014co} to make VMs' deployment unpredictable to attackers. Han et al. have proposed a co-resident attack resistant VM allocation policy \cite{han2017using}, which distributes VMs by optimizing security, workload balance and power consumption needs of cloud servers. Li and Zhang et al. have designed a Vickrey-Clarke-Groves (VCG) mechanism to migrate VMs periodically, so that malicious VMs cannot stay co-located with their target VM for a long time even if they can achieve co-residence \cite{li2012improving}. Chhabra et al.  proposed an allocation policy to reduces the probability of co-residence by classifying legal VMs and attacker VMs based on historical data, similar approaches often require significant computational analysis and previous knowledge on each incoming request~\cite{chhabra2020secure}, which can be further improved.

\vspace{-2mm}
\subsection{\bf Energy-Efficient Cloud Computing}
Besides security, energy-efficient cloud computing has recently attracted great attention as data centers consume a large amount of electricity and generate giant power bills every year at companies like Google, Facebook, Amazon, etc. Data centers consumed more than 2\% of the US total electricity consumption~\cite{EEClouds:GoogleFacebookPower}. Different energy-efficient solutions have been applied at ventilation, liquid-cooling systems, and building construction~\cite{US6574104B2}. However, such construction level modification will generate a large amount of cost. Furthermore, cooling systems will also consume a significant amount of electricity. Without conducting hardware level modification, a power-aware VM scheduling algorithm could significantly reduce energy consumption with minimum financial cost and little performance impact. Recent research shows that VM scheduling algorithms have great impact on overall energy consumption of a data center~\cite{EEClouds:CloudSimConsolidation}. 
Therefore, energy-efficiency is used as an important factor for our scheduling algorithm to evaluate the overall performance and efficiency.


\vspace{-2mm}
\subsection{\bf Ant Colony Algorithm and Its Applications}

The Ant Colony Optimization (ACO) is a meta-heuristic algorithm for finding optimized solutions of computational problems. It is inspired by one behavior of ants, in which they leave pheromone on favorable paths for other members to follow \cite{dorigo2006ant}. ACO has been applied to a wide range of optimization problems which are mostly NP-hard. With the initial application to the Traveling Salesman Problem (TSP)\cite{dorigo1996ant}, ACO has also been applied to solve other problems like sequential order problem (SOP) \cite{gambardella2000ant}, vehicle routing problem \cite{gambardella1999multiple, bell2004ant}, resource constraint project scheduling problem \cite{merkle2002ant}. 

In cloud computing, ant colony optimization is widely used in task scheduling \cite{ragmani2018performed, ragmani2019faco}. Li proposed a Load Balancing Ant Colony Optimization (LBACO) algorithm to achieve task scheduling in dynamic cloud system while in consideration of load balancing at the same time \cite{li2011cloud}. Feller has applied ACO in workload placement and the results show that this approach provides superior energy efficiency~\cite{feller2011energy}. Similar applications can also be seen in \cite{natarajan2021task} and \cite{ming2018multi} where ACO has been adopted to address cloud scheduling tasks. However, they do not consider the security aspect. 

To the best of our knowledge, this is the first work to apply ACO to address the secure VM allocation issue in cloud. Based on its high efficiency and effectiveness in addressing NP-hard problems, we believe ACO is an appropriate tool to allocate cloud VMs so that the cloud's overall security, power consumption and workload balance are optimized. 
\vspace{-2mm}
\subsection{\bf Our Earlier Work}

In~\cite{icnc2018aco}, which is the conference version of the work, we proposed the optimized energy-efficient and security-aware VM allocation strategy against co-residence attack. The preliminary results indicated that the presented research is able to achieve the balance among cloud security, energy-efficiency, and workload balance. The journal version is significantly different from our conference version in the following aspects. {\bf First}, from the model aspect, rather than assuming all VMs arriving at the same time, this work considers a more realistic real-time scenario as a random number of VMs from different users arriving at the cloud with random timings, which requires the solution to dynamically evolve according to the existing VM allocation status and the incoming new VM requests. {\bf Second}, from the solution aspect, to balance computational complexity, real time delay and the optimization results, we first introduce time windows to handle VMs in clusters and then apply ACO algorithm for each time window to manage VM allocation. A more in-depth understanding of the ACO algorithm, how and why it is mapped to address the proposed problem have been discussed in a more comprehensive way, which well explained the fundamental working mechanisms of the proposed scheme. {\bf Third}, as a proof of concept, the conference version only provided basic performance evaluations. More sophisticated experiments and data analysis based on real world cloud workload traces have been conducted in this journal draft. Each of the key parameters of the proposed scheme has been tested and discussed. Additional state-of-the-art comparison scheme has been implemented and compared with the proposed scheme. The results are discussed in details. {\bf Last but not least}, more comprehensive reviews and analysis of the state-of-the-art literature have been conducted.

\section{Modeling}

In this section, we will present the proposed secure VM allocation strategy in details. In particular, we would like to first discuss the system model and assumptions; then model the secure allocation issue as an optimization problem; and present how to adopt ACO algorithm to solve the optimization problem in an efficient way. 

\vspace{-2mm}
\subsection{\bf Assumptions}
As one of the first few works to systematically model the secure and energy-efficient VM allocation problem at IaaS level in cloud, we propose to make the following assumptions to facilitate the establishment of the optimization model later. 

First, we assume the cloud receives a random number of VMs from different users at random timings. Periodically, the cloud needs to assign $n_v^t$ VMs from $n_u^t$ users arriving in the time duration $t$ to a number of available PMs, so that the VM assignment can minimize security risks, overall power consumption and imbalance of workload among PMs. How frequently the cloud should perform such assignment can be determined to balance time delay, computational complexity and optimal results. 

Second, for each time window $t$, the number of PMs involved in the allocation, marked as $n_s^t$, is not given. As we assume that there are sufficient number of idle PMs to host VMs, $n_s^t$ should be a value within the range $[n^t_{s\_{min}}, n^t_{s\_{max}}]$. In particular, the minimum number of PMs, $n^t_{s\_{min}}$, is achieved when all the VMs are squeezed into the minimal number of PMs to make the utilization as high as possible. On the other hand, the maximum number of PMs, $n^t_{s\_{max}}$, is achieved when each VM is assigned to a different PM. In other words, $n^t_{s\_{max}} = n^t_v$. This allocation achieves maximum security since all VMs are isolated on different PMs at the cost of highest power consumption and workload imbalance.  

Third, we assume each VM's workload is dynamically changing during run time based on the real world cloud workload traces. Please note that such changes will lead to fluctuations of the power consumption and workload balance, and may occasionally cause overload of PMs which triggers dynamic VM migrations among PMs in cloud. These above assumptions make our model more realistic but also more challenging to address.

Fourth, regarding the security aspect, we assume that all VMs from a malicious user are malicious. The attack goal is to have the malicious VMs achieve co-residence with VMs from as many normal users as possible to facilitate later attacks. In addition, from the defender side, we also assume that according to historical data, the cloud is able to estimate the percentage of malicious users, but does not know which specific users are malicious. This assumption requires that the proposed scheme can develop the best allocation strategy based on different security context. In the case where the cloud does not have a good estimation of the malicious user percentage, this value can always be set as 100\% to treat security in the most conservative way, which will result in an allocation solution that minimizes the co-locations of VMs from different users.

\vspace{-2mm}
\subsection{\bf Optimization Model} \label{sec:OM}
With the above assumptions, we model the overall VM allocation problem as an optimization problem, of which the optimization goal is to minimize (1) the security risks, which is modeled as the probability of malicious VMs co-locating with the VMs from normal users (i.e. $R_{sec}$), (2) the overall power consumption of used PMs to run these VMs (i.e. $f_{Power}(u)$), and (3) the workload inequality among different PMs (i.e. $B_{w}$). Therefore, we design the objective function of this optimization problem as follows.  

\begin{equation}
\label{eq:overall_cost}
c=w_{1}*R_{sec}+w_{2}*f_{power}(u)+w_{3}*B_{w}
\end{equation}
where $c$ represents the objective function value or cost value, $w_1, w_2,$ and $w_3$ represent weights for the three factors respectively. 

Then the next question is how to quantify these three factors in a reasonable way. In this work, we propose the quantification of these factors as follows.  


\subsubsection{\bf Quantification of Security} Specifically, as we assume a uniform distribution of malicious users, we model the probability of malicious VMs co-locating with normal users (i.e. the security risks) as
\begin{equation}
\label{eq:security}
R_{sec}=P_{mal}*\frac{\sum_{i=1}^{n_s}(n^i_{co-loc} -1)}{n_s*(n_u-1)}
\end{equation}
where $P_{mal}$ indicates the estimated malicious user percentage; and $n_s$ and $n_u$ represent the number of PMs and users, respectively. $n^i_{co-loc}$ indicates the number of co-located users at PM $i$. We can see that (1) in the ideal case, where each PM hosts no more than one user's VMs, $R_{sec}$ is $0$; (2) in the worst case, where each PM hosts VMs from all the users, $R_{sec}$ is $1$; and (3) $R_{sec}$ will increase when either the percentage of malicious users or the number of co-located users at each PM increases.

\subsubsection{\bf Quantification of Power Consumption} The power cost evaluation is based on the power measurement of a PM at eleven CPU utilization levels at 0\%, 10\%, 20\% ... 100\%~\cite{EEClouds:DellPowerEdgeR820} since only CPU utilization is sufficient to evaluation whole computing system's power cost~\cite{GooglePower}. Since measuring power cost at all utilization levels are neither cost-effective nor practical, we use linear interpolation (Eq.~\ref{eq:powerInterpolation}) to estimate the corresponding power cost that is at an unmeasured utilization level $U$ by using the measured power costs $P_h$ and $P_l$ at the higher utilization $U_h$ and lower utilization $U_l$.
\begin{equation}
\label{eq:powerInterpolation}
f_{Power}(u) = \frac{P_h-P_l}{U_h-U_l}U - \frac{P_hU_l-P_lU_h}{U_h-U_l}
\end{equation}

The power cost is normalized as
\begin{equation}
\label{eq:power}
P_{normalized} = \frac{\sum_{i=1}^{n_{s}}{\left | P_i-P_{best} \right |}}{P_{best} \times n_{s}}
\end{equation}
where $P_i$ represents the current power cost of the $i$th PM and $P_{best}$ represents the most effective power cost~\cite{EEClouds:DellPowerEdgeR820} that has the highest performance to power ratio. We can see that the greater the difference between $P_i$ and $P_{best}$, the less power efficient the current server is.

\subsubsection{\bf Quantification of Workload Inequality} At the end, the cost of workload inequality is normalized as
\begin{equation}
\label{eq:workload}
B_{w}=\frac{1}{n_s}\sqrt{\sum_{i = 1}^{n_s}(wl_{i}-\overline{wl})^{2}}
\end{equation}
where $wl_i$ represents the workload of VM $i$; and $\overline{wl}$ represents the average workload for all the $n_s$ PMs. We can see that either an extremely large or extremely small workload will dramatically increase the cost of the workload inequality.

\vspace{-2mm}
\subsection{\bf Dynamic VMs in Real Time}

As proved in other existing studies \cite{han2013security,han2017using}, we recognize the optimization issue modeled in the above section as an NP-complete problem. However, the problem is even more complex as in reality, cloud servers continuously receive dynamic VM requests and the workload of existing VMs is also dynamically changing. 

To make the model more realistic, we assume that users' requests arrive at the cloud end with random timings. Furthermore, the request from each user may be realized through a random number of VMs. In particular, we 1) adopt Poisson distribution to simulate the incoming VMs' arrival rate; 2) introduce time window concept to group incoming requests to balance optimal assignment solution, computational complexity and allocation delay; 3) use real world cloud traces for VM workload simulations.

\subsubsection{\bf Arrival Timing}\label{sec:AT} We adopt Poisson distribution, a discrete probability distribution that expresses the probability of a given number of events occurring in a fixed interval of time, to simulate the arrival time of VMs. Furthermore, the number of VMs arrives in one time interval does not affect that in any other time intervals. 

In particular, VM arrival rate $\lambda$ is used to tune the workload by varying the inter arrival time based on Poisson Distribution as presented in the following equation where $R$ is a random number between 0 and 1, $e$ is the base of natural logarithm, $\lambda$ is greater than 0. Greater $\lambda$ means smaller inter-arrival time between VMs which results in more intense VMs in the same period of time. Therefore, parameter $\lambda$ tunes the real-world workloads to better evaluate our strategies performance. 

\begin{equation}
Interval Time=-log_{e}(R/\lambda). 
\end{equation}

\subsubsection{\bf Time Window} Since VMs are continuously arriving, we propose to handle VM requests in groups through time windows. VMs arriving in the same time window will be processed together. Such solution requires a careful design of the time window length. The search of an optimal allocation strategy can have more flexibility when more VMs are available, which may lead to better performance in achieving the optimization goals. However, waiting to gather too many VMs will cause significant delays to handle users' requests. Meanwhile, with more number of VMs handled together, the computational complexity will also increase, leading to further delays. Therefore, there is a trade-off between request delay and the optimization performance of the resulted allocation strategy. 

\subsubsection{\bf Data Trace} We simulate the workload of each VM based on real world cloud workload traces. As a result, the utilization request of a VM may change from time to time, which requires the VM allocation algorithm to dynamically evaluate the workload at each PM accordingly and migrate some allocated VMs in case of server overloading.

\vspace{-2mm}
\subsection{\bf Ant Colony Optimization}

Next, we propose to adopt Ant Colony Optimization (ACO) as a solution to the proposed optimization model. Inspired by natural ant activities, ACO is an algorithm integrating both heuristic information and randomness to find the optimized solution to a problem~\cite{dorigo2006ant}. The basic idea is that ants carry back their food to colony through different random trails initially, and meanwhile release pheromone on their trails. After a while, trails that take ants less time to travel are piled up with pheromone and become more attractive to ants traveling later. In this way, the shorter trails are reinforced again and again, so that eventually the shortest one will stand out. 

A typical application of the ACO algorithm is traveling salesman problem (TSP), an NP-hard problem, of which the optimization goal is to find the shortest path to traverse all cities in a given map. In particular, given an ant $k$ at a city $i$, the probability for this ant to choose the next city as $j$ is calculated as follows.
\begin{equation}
\label{eq:ant_probability}
p_{ij}^k =  \begin{cases}
			\frac{\tau_{ij}^\alpha *\eta_{ij}^\beta}{\sum_{l \in {\bf N}} \tau_{il}^\alpha *\eta_{il}^\beta} & j \in {\bf N}\\
			0  & \text{otherwise}\\
			\end{cases} 
\end{equation}
where ${\bf N}$ is the set containing all the cities that are not visited by ant $k$ yet; $\tau_{ij}$ and $\eta_{ij}$ represent the pheromone and heuristic value of selecting city $j$ to visit next after city $i$. The heuristic value $\eta_{ij}$ can be calculated based on the direct distance between cities $i$ and $j$ (i.e. $d_{ij}$) as follows. 

\begin{equation}
\label{eq:ant_heuristic}
\eta_{ij} = \frac{1}{d_{ij}}
\end{equation}
\vspace{-2mm}


From the above discussion, we can see that the heuristic value $\eta_{ij}$ is determined by the direct distance between cities $i$ and $j$, which is consistent with the intuitive way of determining the shortest path. In addition, the pheromone value $\tau_{ij}$ may initially represent randomness, as ants may take random trails and lay pheromone, and is gradually reinforced  by later ants' choices/experiences (i.e. if the path from city $i$ to city $j$  is frequently selected as part of the best path). By adjusting the two parameters $\alpha$ and $\beta$, which range from 0 to 1, ACO can dynamically adjust how important the pheromone and heuristic values are considered, respectively. For example, when $\alpha=0$, the pheromone information is completely ignored; and the ACO algorithm becomes a pure greedy algorithm. On the other hand, when $\beta = 0$, the heuristic information is completely ignored; and the ACO algorithm becomes a pure random searching algorithm. 
  
By taking both information into account, the ACO algorithm aims to identify the optimal trail by integrating ``exploitation" (selecting ``optimal" action based on heuristic information that is already known) and ``exploration" (attempting to discover new possibilities by selecting a sub-optimal action with certain randomness). In addition, as different ants are independent from one another, the ACO algorithm can be naturally implemented in a distributed way to improve efficiency. Therefore, the ACO algorithm is often applied on NP-hard problems to efficiently find high quality solutions.

\vspace{-2mm}
\subsection{\bf Adopting Ant Colony Optimization (ACO)}
The advantages of ACO make it a promising approach to address the secure VM allocation issue in cloud. Therefore, we would like to adopt ACO in the proposed scheme to address the optimization problem discussed in Section \ref{sec:OM}. However, such adoption is not trivial due to several challenges. First, how to map the VM assignment issue, which involves two parties as the VM and the PM, to a city visit problem that only considers one party (i.e. cities)? Second, different from TSP where the number of cities to visit is fixed, the secure VM allocation problem only specifies the number of VMs to assign, while leaving the number of physical servers open. How to determine the optimal number of PMs involved? Third, how to model the heuristic and pheromone values in the VM assignment scenario? 

We aim to address these challenges in the following two sections. In particular, we need to handle two major steps as: (1) mapping VM allocation as a shortest path problem, and (2) designing
heuristic and pheromone values in VM allocation.

\subsubsection{\bf Mapping VM Allocation}
We propose to address the first two challenges through the following mapping scheme. Recall that the original VM allocation problem is to assign a list of $n_v$ VMs (i.e. $V_{list}$) to $n_s$ available physical servers, where $n_s \in [n_s^{min}, n_s^{max}]$ is not a fixed value (i.e. the second challenge mentioned above). To simplify the problem, we first divide the entire problem into $n_s^{max} - n_s^{min}$ subproblems, where each subproblem only handles one specific PM number. We will retrieve the optimal solution to the overall problem as the best solution out of the optimal solutions to each subproblem. 

Then for each subproblem with a fixed number of PMs, represented by $n_s$, we aim to assign each VM in the $V_{list}$ to these $n_s$ PMs one by one. Specifically, we create a VM assignment vector $A$ as 
\begin{equation}
A^{n_s} = [a_0, a_1, ... a_i, ... a_{n_v-1}]
\end{equation}
where $a_i$ represents the PM index that the $i^{th}$ VM in the $V_{list}$ is assigned to. For example, given an assignment $A^3 = [1, 0, 2, 1]$, it indicates that four VMs have been assigned to three different PM as server 1, server 0, server 2 and server 1, respectively. The first challenge mentioned above can then be addressed through this VM assignment vector $A$. Similar to the TSP problem, where each traversal solution contains a specific order of cities that leads to a certain overall distance; in our problem, each VM assignment $A$ contains a specific combination of VM-PM matching pairs that leads to a certain overall cost.


\subsubsection{\bf Heuristic and Pheromone Information in VM Allocation}
Here, we address the third challenge: determining the heuristic and pheromone values in the VM allocation problem. Recall that in TSP, the heuristic value is determined as the inverse of the distance between two cities. In our problem, we design the heuristic value $\eta_{ij}$ as a value related with the cost of assigning VMs to PMs. To record all the assignment costs, we introduce a two dimensional cost matrix ${\bf C}$ as

\begin{equation}
{\bf C} = 
\begin{bmatrix}
    c_{1,1}       & c_{1,2} & c_{1,3} & \dots & c_{1,n_s} \\
    \hdotsfor{2} & c_{\nu,j} & \hdotsfor{2} \\
    \hdotsfor{5} \\
    c_{n_v,1}       & c_{n_v,2} & c_{n_v,3} & \dots & c_{n_v,n_s}
\end{bmatrix}
\end{equation}
where $c_{\nu,j}$ represents the cost of assigning VM $\nu$ to server $j$, which is calculated according to equation (\ref{eq:overall_cost}), as the extra cost increase on security risks, power consumption, and workload inequality caused by the assignment of VM $\nu$ to PM $j$. Then the heuristic information $\eta_{ij}$ can be easily obtained from this matrix as

\begin{equation}
\eta_{ij} = \frac{1}{c_{\nu,j}}
\end{equation}
Please note that as we assign VMs according to their orders in the $V_{list}$, the cost $c_{\nu,j}$ is calculated based on the previous assignment of VM $\nu-1$ to server $i$ (i.e. $c_{\nu-1,i}$). Therefore, different orders of the VMs in the $V_{list}$ may lead to different assignment solutions.

Second, we design a two-dimension pheromone matrix ${\bf Ph}$ as follows.
\begin{equation}
{\bf Ph} = 
\begin{bmatrix}
    ph_{1,1}       & ph_{1,2} & ph_{1,3} & \dots & ph_{1,n_s} \\
    \hdotsfor{2} & ph_{\nu,j} & \hdotsfor{2} \\
    \hdotsfor{5} \\
    ph_{n_v,1}       & ph_{n_v,2} & ph_{n_v,3} & \dots & ph_{n_v,n_s}
\end{bmatrix}
\end{equation}
where $ph_{\nu,j}$ represents the current pheromone value of assigning VM $\nu$ to server $j$. In the beginning of the problem, as there is no information available for possible assignments, all the values in the initial pheromone matrix are normalized as $\frac{1}{n_s}$. Once a local optimal assignment has been identified, the VM-PM matching pairs that are involved in this assignment will have their pheromone values updated.

\subsubsection{\bf Iterative ACO for Secure VM Allocation}
With all the three challenges addressed, we are now able to present the iterative ACO algorithm for secure VM allocation. 

Step 1, divide the original problem of ``assigning a list of $n_v$ VMs to $n_s$ PMs, where $n_s \in [n_s^{min}, n_s^{max}]$" into $n_s^{max} - n_s^{min}$ subproblems. For each subproblem with a fixed number (i.e. $n_s$) of PMs, an iterative ACO will be launched to find the optimal assignment. 

Step 2, for each iteration $l$, identify the best assignment. Specifically, $n_a$ ants are created, where each ant will start from a $V_{list}$ with a randomly generated order of VMs, and work on constructing its own assignment $A^{n_s,l}$ by considering both the heuristic and pheromone information. Once all the $n_a$ ants have completed their assignments, the total cost of each assignment is stored as an element in a $1 \times n_a$ vector $C^{t,l}$. The assignment with the lowest cost $min(C^{t,l})$ will be identified as the best assignment (i.e. $A_{opt}^{n_s,l}$) for iteration $l$. 

Step 3, update information. If the minimum cost at iteration $l$ is smaller than the optimal cost for the current subproblem (i.e. $min(C^{t,l})< c^{n_s}_{opt}$), we will have 

\begin{equation}
c_{opt} = min(C^{t,l}) 
\end{equation}

\begin{equation}
A_{opt}^{n_s} = A_{opt}^{n_s,l}
\end{equation} 
Consequently, the pheromone information for the next iteration $l+1$ will be updated as follows.

\begin{equation}
ph_{\nu,j}^{(l+1)} = (1-\varphi)ph_{\nu,j}^{l} + \varphi \Delta ph_{\nu,j}^l,
\end{equation}
where $\Delta ph_{\nu,j}^l= \frac{1}{c^{n_s}_{opt}}$. In addition, $\varphi$ represents how fast the pheromone information is updated. A higher $\varphi$ value represents a faster speed to forget the out-of-date pheromone information.

Step 4, repeat the above process for $L$ iterations in total. In particular, each new iteration will be performed by another $n_a$ ants with the updated pheromone information, which may lead to some new better assignments. After $L$ iterations, the final $A_{opt}^{n_s}$ will be determined as the best solution for this subproblem.

Step 5, once all the subproblms are addressed, the global optimal assignment \begin{equation}
A_{opt}^{Global} = \text{optimal}(A_{opt}^{n_s}), \text{ where } n_s \in [n_s^{min}, n_s^{max}]
\end{equation}

We summarize the proposed scheme in Algorithm~\ref{algrithm:VMAllocation}. For clarity purpose, we use bold notations to represent matrices, capital notations to represent vectors and lower case notations to represent scalar variables. The time complexity of Algorithm~\ref{algrithm:VMAllocation} is affected by the max and min number of servers, the number of iterations, the number of ants, and the number of VMs. Since the numbers of iterations and ants are fixed for each execution, the time complexity is roughly $O(n^2)$. 
\begin{table}
  \centering
  \caption{Table of notations} \label{tab:notations}
  \begin{tabular}{|c|c|}
    \hline
	Notation & Description\\ \hline
    $U_{list}$ & A list with all users\\ 
    $V_{list}$ & A list with all VMs\\ 
    $n_s$ & Number of servers \\
    $n_u$ & Number of users \\
    $n_v$ & Number of VMs \\
    $n_a$ & Number of ants \\
    $L$ & Number of iterations \\
    ${\bf Ph}$ & Pheromone Matrix\\
    ${\bf C}$ & Cost matrix for assigning VMs to different servers\\
    $C_a$ & A vector of costs for each ant's assignment\\
    $A$ & VM assignment vector with dimension as $n_v*1$ \\
    $A_{opt}^{n_s}$ & optimal VM assignment vector for $n_s$ servers\\
    $A_{opt}^{Global}$ & The global optimal VM assignment \\
    \hline
  \end{tabular}
\end{table}

\begin{algorithm}
	\centering
	\caption{ACO Cloud VM Assignment Algorithm}
	\label{algrithm:VMAllocation}
	\begin{algorithmic}
		\STATE $U_{list}$ $\leftarrow$ All Users
		\STATE $V_{list}$ $\leftarrow$ All VMs 
		\STATE $n_s^{min}, n_s^{max}$ $\leftarrow$ The max/min number of servers
		
		\FOR {$n_s = n_s^{min}$ to $n_s =n_s^{max}$}
			\STATE Initialize pheromone matrix ${\bf Ph}$, $c^{n_s}_{opt} = Inf$ and $A_{opt}^{n_s} = NULL$
			\FOR {$l=0$ to $l=L-1$}
				\FOR{$m = 0$ to $m= n_a-1$}
					\FOR{$\nu=0$ to $\nu= n_v -1$}
						\STATE ${\bf C}_v[\nu] = getCost(\nu, V_{list})$
						\STATE ${\bf Pr}_v[\nu] = getPro({\bf C}_v[\nu], {\bf Ph}, \alpha, \beta)$
						\STATE $A^{n_s,l}[q] = randomGen({\bf Pr}_v[\nu])$
					\ENDFOR
					\STATE $C_a^l[m] = \sum_{\nu=1}^{n_v} C_v[\nu][A^{n_s,l}[\nu]]$
					\IF {$c^{n_s}_{opt} >C_a^l[m]$}
						\STATE $c^{n_s}_{opt} = C_a^l[m]$ and $A_{opt}^{n_s} = A^{n_s,l}_{opt}$
					\ENDIF
				\ENDFOR
				\STATE ${\bf Ph} = pheUpdate(c^{n_s}_{opt})$
			\ENDFOR
		\ENDFOR
		\STATE return $A_{opt}^{Global} = optimal(A_{opt}^{n_s})$, where $n_s \in [n_s^{min}, n_s^{max}]$
	\end{algorithmic}
\end{algorithm}

\section{Performance Evaluation}

To demonstrate the effectiveness and efficiency of the presented solution, we evaluate our strategy under different amount of users and workloads. Then we compare the performance with two other state-of-the-art VM allocation strategies. The results of simulated experiments indicate that the proposed strategy outperforms the baseline strategies considering workload balance, security, and power cost. Table~\ref{EECloudSim:VM} presents the standard VM configuration that is used in the experiments. Bandwidth and VM size decide the migration time cost. MIPS (Millions of Instructions per Second), Processing Elements, and VM utilization will be used to convert the VM utilization to server utilization. In particular, the real time VM utilization rate traces are generated based on data collected from a real data center. Table~\ref{EECloudSim:PPRofGears} presents the power and performance data of the servers used in the simulation.

\begin{table}
	\caption{Virtual Machine Configuration}
	\centering
	\begin{tabular}{|c|c|}
		\hline
		\label{EECloudSim:VM}
		Configuration Parameters & Default Value\\
		\hline \hline
		MIPS & 2000\\
		Processing Element & 2 \\
		Memory& 1GB\\
		Bandwidth & 100 Mbit/s\\
		VM Size& 2.5 GB\\ 
		VM CPU Utilization & 0\% - 100\%\\
		\hline
	\end{tabular}
\end{table}

\begin{table}
	\caption{Power Consumption of Hosts~\cite{EEClouds:FUJITSUPRIMERGYRX1330}~\cite{EEClouds:InspurNF5280M4}~\cite{EEClouds:DellPowerEdgeR820}~\cite{EEClouds:IBMNeXtScalenx360}}
    
\hskip-2.0cm\begin{tabular}{|c|c|c|c|c|c|c|c|c|c|c|c|}
		\hline
		\label{EECloudSim:PPRofGears}
		Host Model & \multicolumn{11}{|c|}{Average Active Power (Watt)} \\
		\hline \hline
		Utilization: & 0\% & 10\% & 20\% & 30\% & 40\% & 50\% & 60\% & 70\% & 80\% & 90\% & 100\%\\
		\hline
		Fujitsu Primergy RX1330 M1 & 13.8 & 20.8 & 23.9 & 26.3 & 29.1 & 32.6 & 36.2 & 42.0 & 48.6 & 55.9 & 63.7\\	
		Inspur NF5280M4 & 44.4 & 83.3 & 101 & 118 & 135 & 146 & 161 & 190 & 218 & 255 & 301\\
		Dell PowerEdge R820 & 71.8 & 135 & 156 & 176 & 198 & 219 & 243 & 269 & 297 & 318 & 374\\
		IBM NeXtScale nx360 M4 & 497 & 814 & 947 & 1079 & 1211 & 1344 & 1493 & 1648 & 1863 & 2108 & 2414\\
		\hline
	\end{tabular}
\end{table}

\vspace{-3mm}
\subsection{\bf Key Parameters Testing}

\subsubsection{\bf ACO Parameters}
In this section, we mainly investigate the impact of three key parameters from the ACO algorithm: $\alpha$ and $\beta$. Specifically, $\alpha$ and $\beta$ are ranging from 0 to 1, representing the weights of the pheromone and the heuristic values to be considered in the optimization, respectively. %

\begin{figure}[!t]
	\centering
	\includegraphics[width=0.55\textwidth]{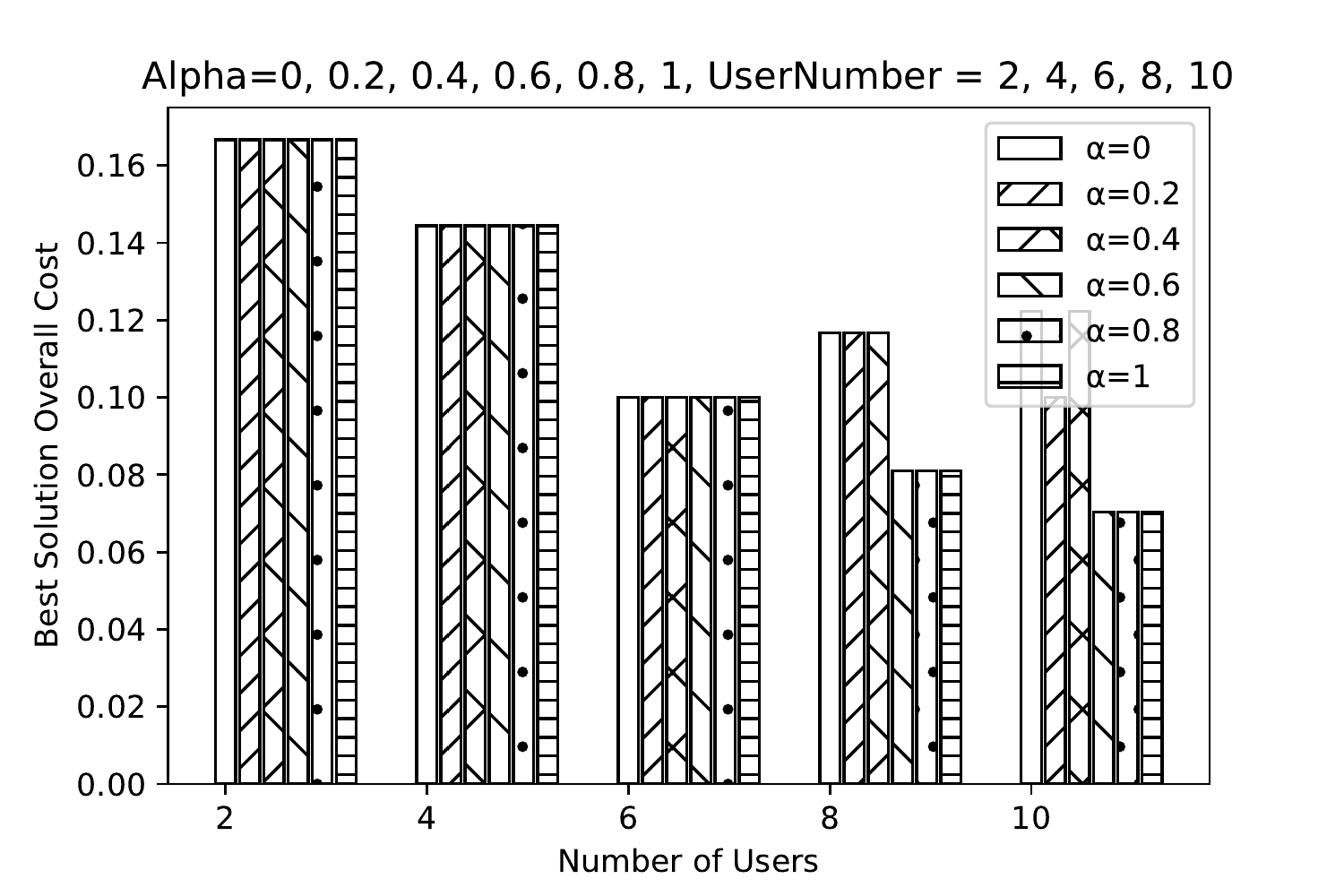}
	\vspace{-2mm}
\caption{Impact of alpha on overall costs}
\label{fig_alpha_impact}
\end{figure}

{\bf Impact of $\alpha$:} To validate the impact of $\alpha$ on the overall costs, we fix $\beta = 0.9$, $\varphi =0.8$ and change $\alpha$ from 0 to 1. The results are shown in Fig.~\ref{fig_alpha_impact}. In particular, we can observe that the overall costs are not sensitive to $\alpha$ values when the number of users is small. When there are more users ($>6$), greater $\alpha$ values will result in smaller overall costs in general. This is reasonable. Recall that the $\alpha$ value indicates the weight of pheromone information to be considered, which starts from an identical value for all possible VM-PM matching pairs and needs to be accumulated over time. When there are not many users/VMs to assign, there is not sufficient accumulation for the pheromone value to represent better matching pairs. As a result, a higher or lower weight (i.e. $\alpha$) of the pheromone values will not influence the overall costs much. However, when more users/VMs are available, the pheromone information can be accumulated more to represent better matching pairs. Therefore, a higher weight (i.e. greater $\alpha$ value) will effectively help to reduce the overall costs.    



\begin{figure}[!t]
	\centering
	\includegraphics[width=0.55\textwidth]{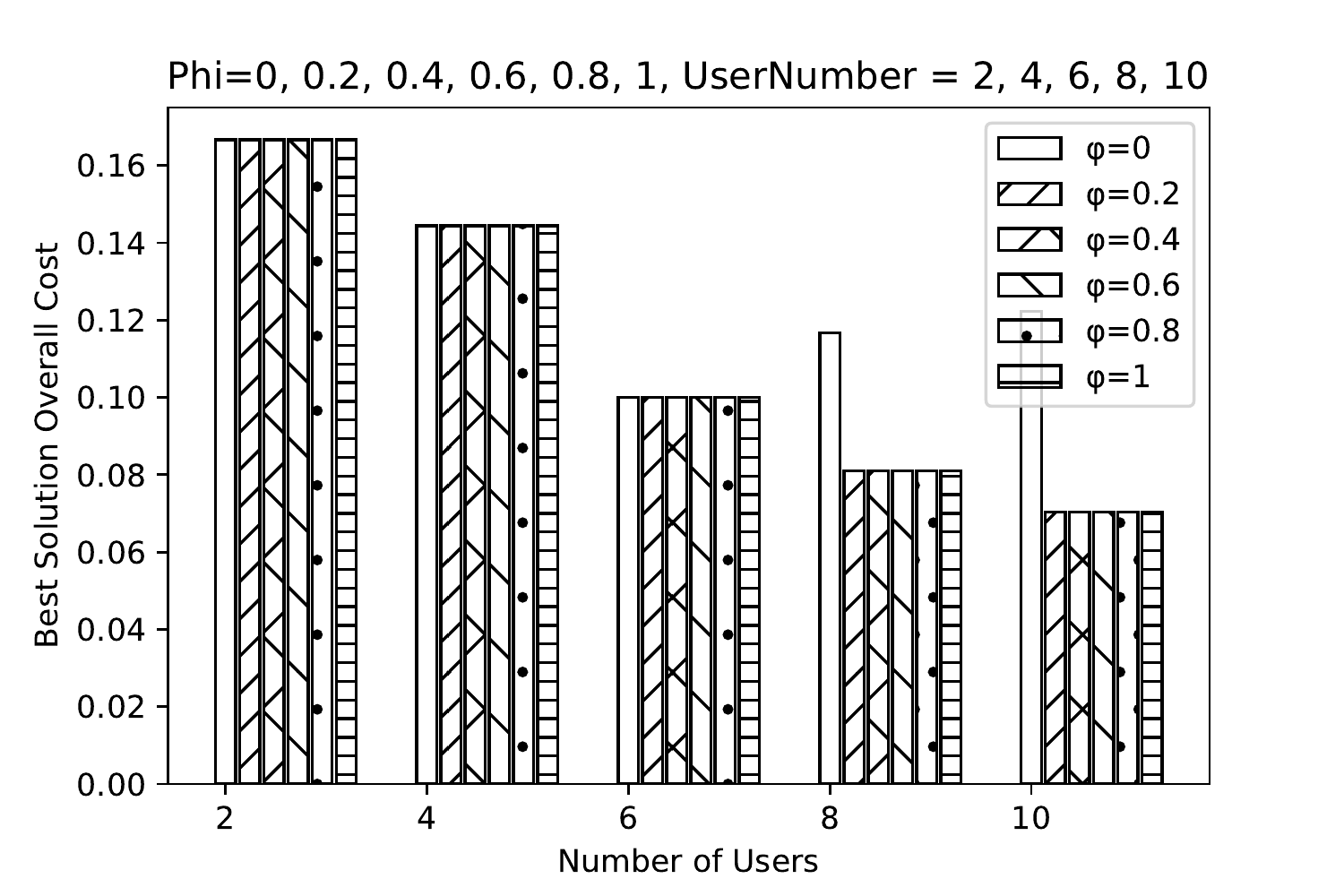}
	\caption{Impact of phi on overall costs}
	\label{fig_phi_impact}
	\vspace{-2mm}
\end{figure}

{\bf Impact of $\varphi$:} Similarly, to evaluate the impact of $\varphi$ on the optimized overall costs, we set $\alpha=0.9$, $\beta = 0.89$ and change $\varphi$ from 0 to 1. The similar trend is observed in Fig.~\ref{fig_phi_impact}, which indicates that although the overall costs can yield lower values when there are more users/VMs, they are not sensitive to the change of $\varphi$ values for a fixed number of users, especially when the user number is small ($<8$). However, when more users/VMs need to be allocated, $\varphi=0$ will not lead to good results. The reason is as follows. Recall that a larger $\varphi$ value indicates a faster update of the pheromone value, $\varphi=0$ leads to no pheromone updates at all. It indicates that the pheromone values in all the later iterations will be exactly the same as the initial pheromone values, which are set as an identical value for all possible VM-PM matching pairs. It actually mitigates the impact of pheromone as it cannot be used to help differentiate different matching pairs. In other words, when only considering heuristic information, the overall cost will not yield its optimal value. But except the case $\varphi=0$, other $\varphi$ values will yield the same optimal overall costs, indicating that regardless of the pheromone updating speed, as long as it is not zero, the optimal overall costs can always be achieved.

%

\begin{figure}[!t]
	\centering
	\includegraphics[width=0.55\textwidth]{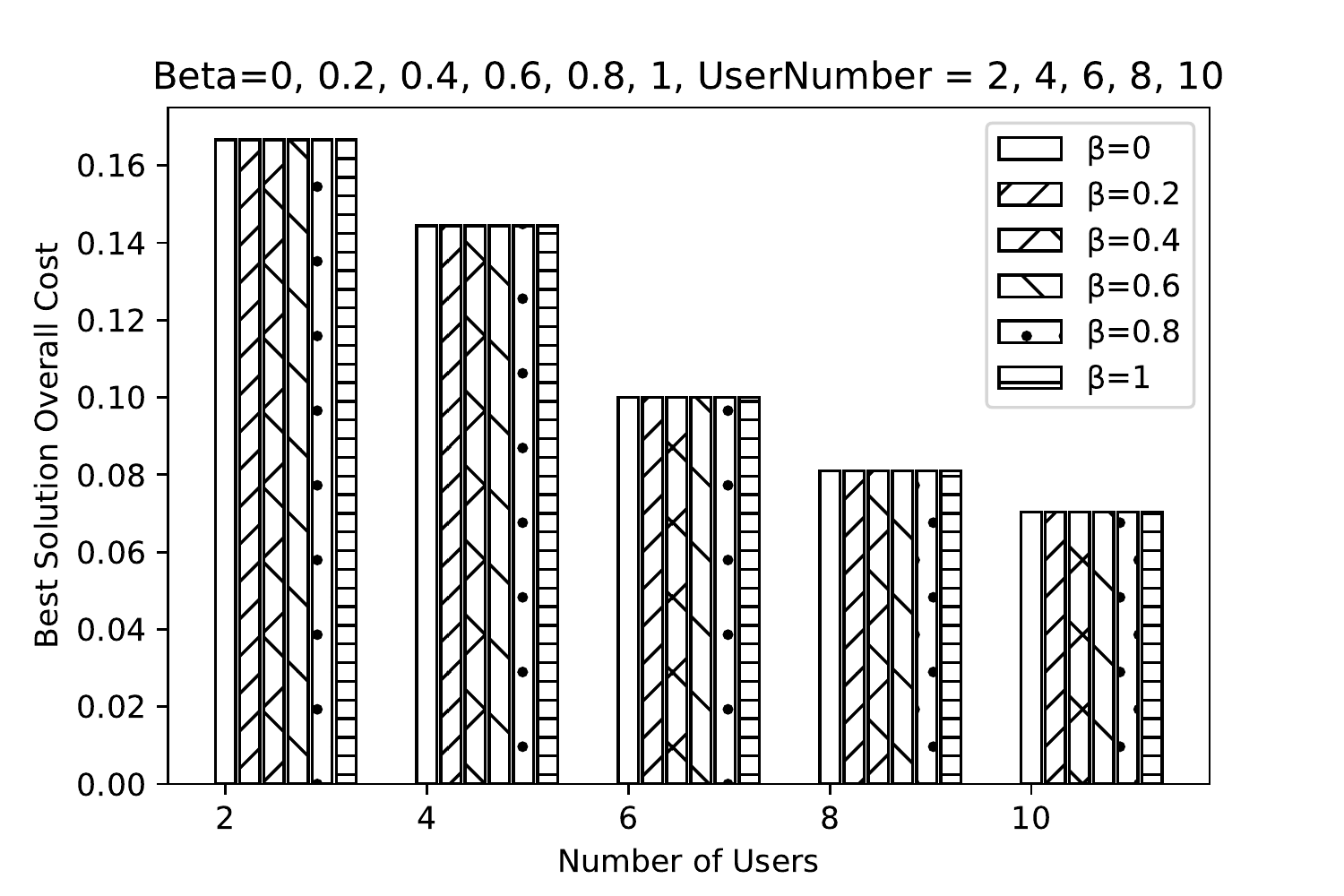}
	\caption{Impact of beta on overall costs}
	\label{fig_beta_impact}
	\vspace{-4mm}
\end{figure}
{\bf Impact of $\beta$:} To validate the impact of $\beta$ on the overall costs, we set $\alpha = 0.9$, $\varphi =0.8$ and change $\beta$ from 0 to 1. The results are shown in Fig.~\ref{fig_beta_impact}. Similar trend can be observed as that although the overall costs can yield lower values when there are more users/VMs (i.e. easier to tune and balance), they are not sensitive to the change of $\beta$ values when the number of servers is fixed. It indicates that a wide range of $\beta$ values can be chosen for the ACO algorithm and will not lead to a dramatic performance change. 

\subsubsection{\bf Percentage of Malicious Users}

\begin{figure}[!t]
	\centering
	\includegraphics[width=0.55\textwidth]{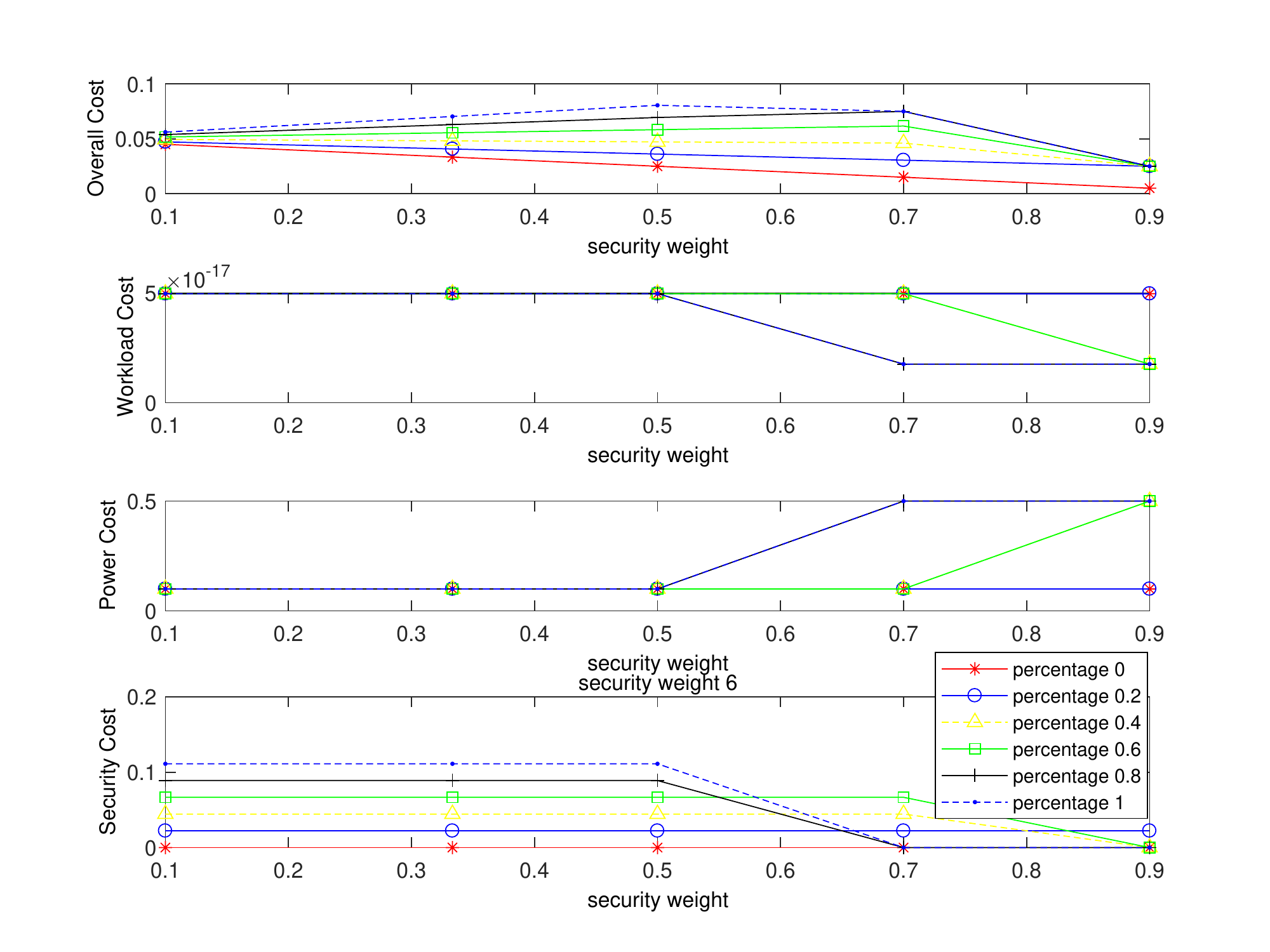}
	\caption{Impact of Malicious User Percentage}
	\label{fig_cost_secweight}
	\vspace{-4mm}
\end{figure}

Besides the ACO parameters, we also want to evaluate how the percentage of malicious users will influence the overall costs. In particular, we have changed the malicious user percentage from 0 to 1, and the security weight from 0.1 to 0.9. The resulted overall costs are shown in Fig.~\ref{fig_cost_secweight}. In Fig.~\ref{fig_cost_secweight}, the x-axis and y-axis represent the malicious user percentage and overall costs, respectively. The five curves represent different weights of the security factor in our objective function (i.e. Equation \ref{eq:overall_cost}). There are several observations. First, when the malicious user percentage increases, smaller security weights often lead to slower increase of the overall costs. This is because when security is less cared (i.e. lower weights), the overall costs are less sensitive to the malicious user percentage changes. Second, when security weights are set high (e.g. 0.7 or 0.9), we can observe that the cost curves achieve their peek values at a certain point and will not continuously increase when the malicious user percentage increases. This is because higher security weights make the optimization process tilted towards the security aspect, which can effectively guarantee no raises of the security risks while malicious user percentage continuously growing. It also shows the effectiveness of the proposed optimization scheme. 

\subsubsection{\bf The Weight of Security}

\begin{figure}[!t]
	\centering
	\includegraphics[width=0.75\textwidth]{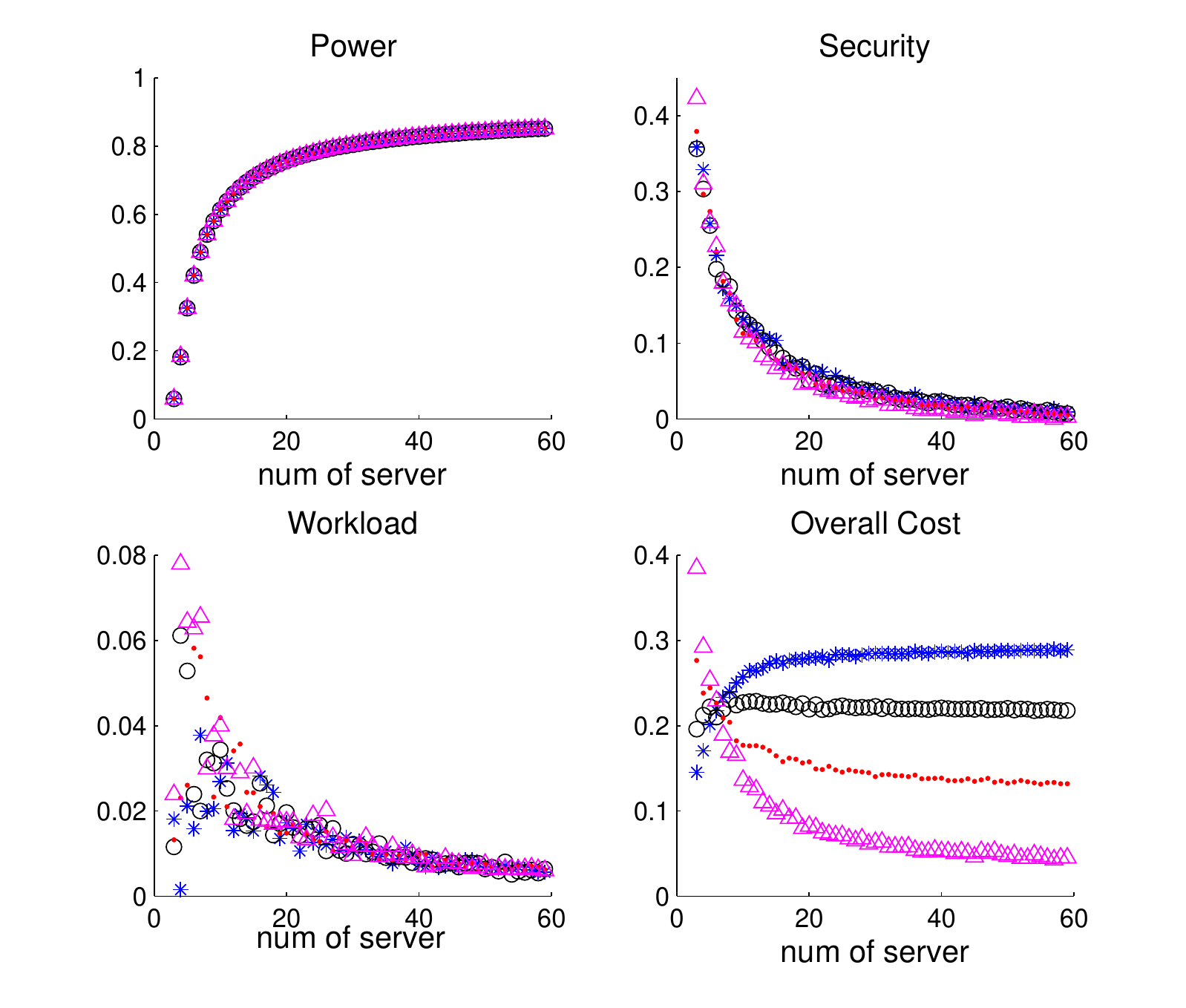}
	\caption{Impact of Security Weights Per Server Num}
	\label{fig_Cost_securityweight_30Users_RandomVMs}
	\vspace{-4mm}
\end{figure}

As the weight of security is also a key parameter to be determined by the cloud service provider, we aim to study its impact on the overall costs, which consists of the costs for power, workload inequality, and security. Specifically, the experiments are conducted by simulating 30 users, where each user has two VMs with random utilization requests. Recall that our proposed algorithm actually divides the entire optimization problem into smaller subproblems, with each one of which handles only a fixed number of servers. So we first present the impact of security weights for each specific number of servers (i.e. each subproblem) in Fig.~\ref{fig_Cost_securityweight_30Users_RandomVMs}. Please note that each of the data points here represents the optimal costs for a specific subproblem, not the global optimal costs.

In particular, there are four subplots, showing the overall costs, workload inequality costs, power costs, and security costs. In each subplot, there are four curves representing different security weights as 0.3, 0.5, 0.7 and 0.9, respectively. The corresponding weights of power consumption and workload inequality are set as $w_{workload balance} = w_{power} = (1-w_{security})/2$. In addition, the x-axis of each subplot represents the number of occupied physical servers, and the y-axis represents the corresponding costs.

From this figure, we can make several observations. First, as shown in the upper left subplot, the power costs are not sensitive to security weights, but mainly dominated by the number of PMs. Even if different security weights lead to different optimal assignment solutions, as long as the solutions are occupying the same amount of PMs, the power costs for these assignments will be roughly the same. 
 
Second, from the upper right subplot, we can observe that regardless of the security weight, when the number of occupied PMs increases, the security costs (i.e. security risks) are decreasing. This is because when the number of occupied PMs increases, the VMs are more spread out, indicating a higher possibility for VMs from different users to be allocated on different PMs, leading to lower security risks/costs. 

Third, from the lower left subplot, we can observe that for all different security weights, the workload inequality costs will always increase first and drop later, when the number of PMs is increasing. The reason is as follows. When the number of PMs is small, most of the VMs are squeezed in the PMs, leading to very limited extra capacity for each PM. Therefore, the workload is roughly balanced among different PMs. When the number of occupied PMs starts to increase, VMs are allocated more flexibly to different PMs, which easily makes more PMs have different capacity left, leading to more imbalanced workload. However, as the number of occupied PMs continues to grow, VMs are spread out, leading to very few VMs sharing the same PM. As a result, it becomes easier again to balance the workload among different PMs.


At the end, as shown in the lower right plot, the overall cost is a trade-off of the three factors: power consumption, workload inequality and security risks, and therefore can be significantly influenced by the security weight. Specifically, when the number of occupied PMs is small, the costs of power and workload inequality can be small. However, as VMs are squeezed to reach the maximum capacity of each server, the security risk is greatly increased. As a result, the local optimal solution will yield higher overall costs if the security factor is the dominated factor (i.e. high security weight), and lower overall costs if power and workload inequality are dominated factors (i.e. low security weight). On the other hand, when the number of occupied PMs is large, the power cost goes up. However, as VMs are spread out on different PMs, the security risks and workload inequality can be low. As a result, the local optimal solution will yield lower overall costs if the security factor is the dominated factor (i.e. high security weight), and higher overall costs if power is the dominated factor (i.e. low security weight).

\begin{figure}[!t]
	\centering
	\includegraphics[width=0.65\textwidth]{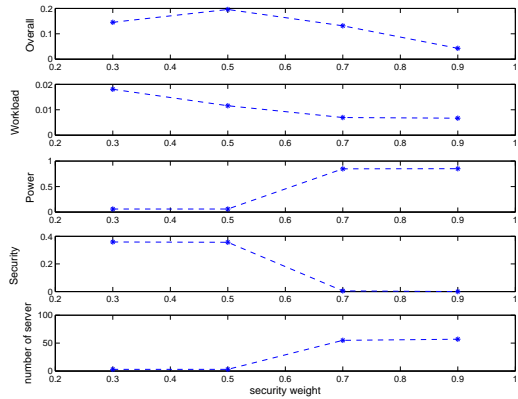}
	\vspace{-4mm}
	\caption{Impact of Security Weights on Optimal Costs}
	\label{fig_optCost_securityweight_30Users_RandomVMs}
	\vspace{-4mm}
\end{figure}

Next, we aim to study the impact of security weights on the global optimal solution in Fig.~\ref{fig_optCost_securityweight_30Users_RandomVMs}. Different from Fig.~\ref{fig_Cost_securityweight_30Users_RandomVMs}, where the local optimal costs for each subproblem are analyzed, here we only examine the global optimal solution with the best number of PMs for each specific security weight. Specifically, the five subplots represent the influence of security weights on the optimal costs for the overall solution, workload inequality, power, and security, as well as the optimal number of servers, respectively. 

From Fig.~\ref{fig_optCost_securityweight_30Users_RandomVMs}, we can observe that when security weight gradually increases, the corresponding optimal solutions tend to make more efforts on lowering the security costs, which will lead to more number of occupied PMs and higher power costs, but lower workload inequality and overall costs.

\subsection{\bf Performance Comparison}

\begin{figure*}[!t]
	\centering
	\includegraphics[width=1\textwidth]{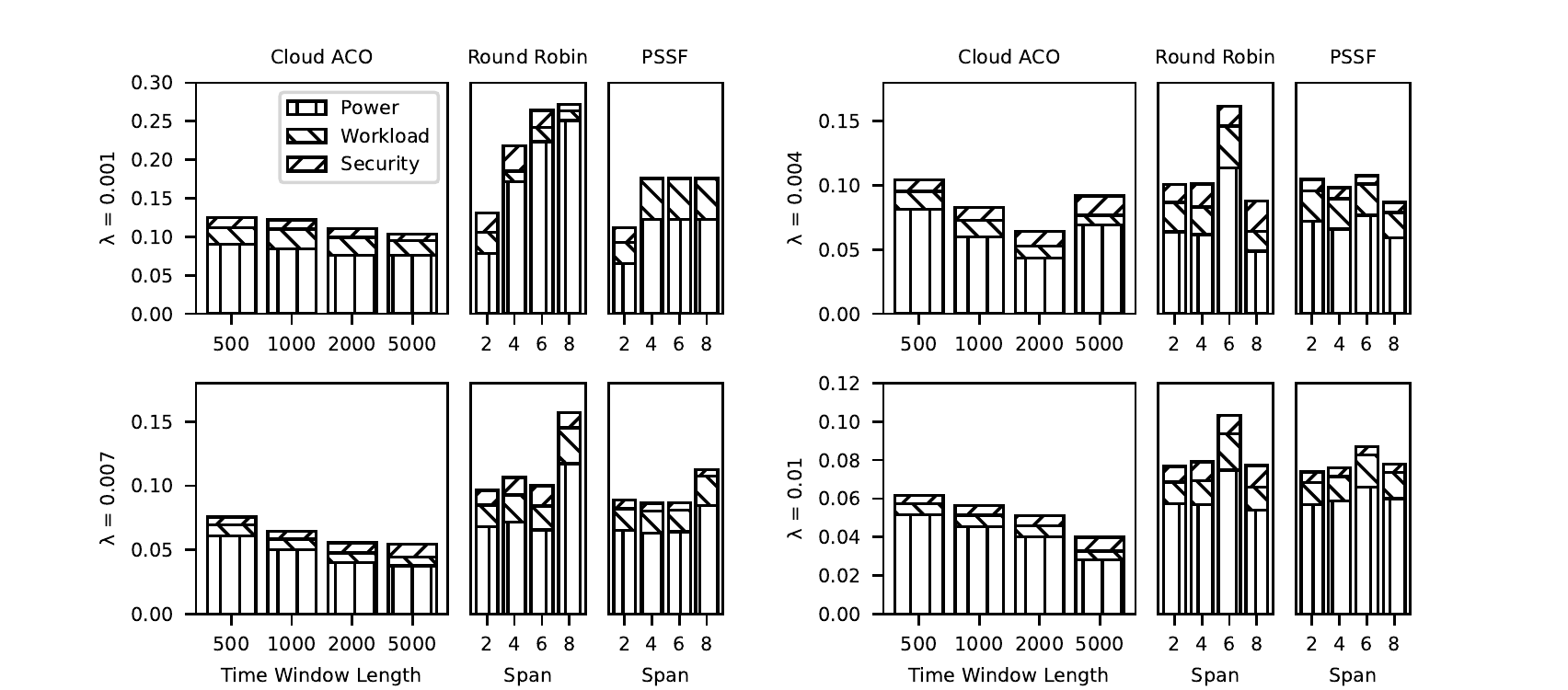}
	\caption[Optional caption for list of figures]{Performance Comparison for Scenario with different $\lambda$}
	\label{SubChannelVMProject:RRACO}
\end{figure*}

In this section, we compare the proposed scheme with two other existing allocation strategies. The first one is Round-Robin, a classic algorithm that allocates resources, e.g. physical machine (PM), in equal portions and in circular order, handling all VMs without priority. In particular, as the original Round-Robin algorithm cannot specify the number of physical PMs to be involved for allocation, we implement the algorithm in a way that the algorithm spawns a group of PMs each time. We set the number of PMs in a group as 2, 4, 6 and 8 each time a new spawning process is needed. The modified Round-Robin algorithms are named RR2, RR4, RR6, RR8 respectively and we examine their performances at different choices of the number of spawning PMs.

The second one is the Previously-Selected-Servers-First (PSSF) scheme proposed in \cite{han2017using}, a representative study of the state-of-the-art VM allocation schemes that optimizes security, workload balance and power consumption through a heuristic strategy. In particular, PSSF tends to select from two strategies: stacking or spreading. Each new VM will be first stacked to the same PM to which other VMs from the same user has been allocated. If the PM has reached its capacity, the new VM will be spread to a new PM. Similar to Round-Robin, PSSF cannot explicitly determine the number of physical PMs to involve. Instead, it involves a group of physical PMs each time, and the new group of PMs will not be involved until the existing PMs reach their capacity. Therefore, a key parameter for PSSF is the number of PMs in each group. In our experiments, we set the number as 2, 4, 6 and 8, respectively to examine its performance and name the algorithms as PSSF2, PSSF4, PSSF6, and PSSF8, respectively.

As discussed in Section \ref{sec:AT}, we use Poisson distribution to simulate different scenarios where the number of users and VMs increases as the parameter $\lambda$ increases in a given amount of time. In particular, $\lambda$ varies from 0.001 to 0.01, which indicates the number of incoming VMs ranging from 10 to 100 during the total experiment duration. We generate 100 sets of data from Poisson distribution at each $\lambda$ values and take the average performance to make a fair comparison. The representative results are presented in Fig.~\ref{SubChannelVMProject:RRACO}.

Fig.~\ref{SubChannelVMProject:RRACO} contains 4 subplots with $\lambda$ set to 0.001, 0.004, 0.007 and 0.01, representing the scenarios with the VM numbers as 10, 40, 70 and 100 respectively to demonstrate the performances of the proposed scheme, Round-Robin and PSSF. In addition, the malicious user percentage is set as 20\%, a reasonable estimation on the malicious context, and the weights for security, power, and workload inequality are all set as 1/3. The bars from left to right represent the proposed scheme with its key parameter (i.e. time window length) equals 500, 1000, 2000 and 5000; Round-Robin with span set to 2, 4, 6, and 8; and lastly PSSF with span set to 2, 4, 6, and 8.

As shown in Fig.~\ref{SubChannelVMProject:RRACO}, the total cost of the proposed scheme on the top right subplot, where $\lambda$ equals 0.004, drops at time window length 200 and then increases along with the increase of time window length. This is because when the incoming number of VMs are relatively small, the number of VMs falling in one time window may vary, leading to some inconsistencies in the performance. As the number of VMs increases where $\lambda$ equals 0.007 and 0.01 on the bottom two subplots, the proposed schemes tends to be more stable with a downward slope as time window length increases. This is because when time window length increases, a larger portion of incoming VMs are considered by the algorithm at each calculation, thus results in a lower overall cost. On the other hand, larger time window length means longer average wait time for each incoming VM before assigning to a actual PM. The trade off here has to be considered in a real world scenario to accomplish an optimized configuration.

Among these algorithms, the proposed scheme always achieves the best performance in terms of the optimal total costs, which validates its effectiveness. Moreover, the detailed subcosts for security, power, and workload balance are also shown for each allocation scheme. There are several observations. First, at $\lambda$=0.001, the proposed scheme with time window length set to 100 achieves the best overall score. PSSF4, PSSF6 and PSSF8 achieves the exact same score with zero cost on security. This is because when the number of incoming VMs is relatively small (10 VMs in this case), according to the behaviour of PSSF, VMs will simply span to a new server when there are enough PMs allocated. This results in zero security cost. However, because of the usage of new PMs, the power costs will significantly increase, leading to much higher overall costs. Secondly, when $\lambda$=0.004, the proposed scheme with time window length 2000 achieves the minimum overall costs. Even though PSSF6 has the lowest security cost, the proposed scheme has far less power cost compared to PSSF6, which results in an overall lowest cost. RR2, RR4, RR6 and RR8 behave similarly to PSSF, but with higher security cost since they tend to stack VMs from different users into the same PM. When $\lambda$=0.007, there are total 70 incoming VMs, which makes it a more realistic scenario. The proposed scheme outperforms both RR and PSSF, and the worst score of the proposed scheme is nearly the same as the best score of both RR and PSSF. Both RR and PSSF have significantly larger power cost and workload balance cost than the proposed scheme because the way the stack up VMs and spawning PMs, and the choices of span 2, 4, 6, 8 of the two schemes among different $\lambda$ does not always generate the same results, which make it hard to detect the optimized choice of span in a complex, realistic scenario. Similar observation can be concluded on the last subplot when $\lambda$ is set to 0.01 with a total of 100 incoming VMs, this confirms that the proposed scheme is robust and scalable with better performance among all there schemes.

\section{Conclusion}

Co-residence attack has raised significant concerns as the increassing popularity of cloud computing. Attackers are able to take advantage of the resource sharing in multi-tenant cloud to perform diverse attacks against their co-residents on the same physical server. We proposed to defend against such co-residence attacks through a secure, workload-balanced, and energy-efficient VM allocation strategy. and modeled the VM allocation problem as an optimization problem. As this optimization problem is NP-hard, we further applied the Ant Colony Optimization (ACO) algorithm, an evolutionary algorithm inspired by natural ant activities, to identify the optimal allocation strategy. Experiment results demonstrated that the proposed scheme can make the multi-tenant cloud secure and power efficient.


%








\bibliographystyle{plain}
\bibliography{ACO_jrnl_compsoc}

\end{document}